\documentclass[communicationpaper,prl,twocolumn,showpacs]{revtex4}
\usepackage{graphicx,bm}
\usepackage{amsmath}
\usepackage{mathrsfs}
\usepackage{color}

\newcommand{\tlv}{{\tilde{v}}}
\newcommand{\tlV}{{\tilde{V}}}

\begin{document}

\title{Particle-wave duality in quantum tunneling of a bright soliton}

\author{Ching-Hao Wang$^{1,2}$, Tzay-Ming Hong$^{1}$ , Ray-Kuang Lee$^{2,3}$, and Daw-Wei Wang$^{1,2,4}$}

\affiliation{$^{1}$Department of Physics, National Tsing-Hua University, Hsinchu 30013, Taiwan
\\
$^{2}$ Frontier Research Center on Fundamental and Applied Sciences of Matters, National Tsing-Hua University, Hsinchu 30013, Taiwan\\
$^{3}$ Institute of Photonics Technologies, National Tsing-Hua University, Hsinchu 30013, Taiwan
\\
$^4$ Physics Division, National Center for Theoretical Sciences, Hsinchu 30013, Taiwan
}

\date{\today}

\begin{abstract}
One of the most fundamental difference between classical and quantum mechanics is observed in the particle tunneling through a localized potential: the former predicts a discontinuous transmission coefficient ($T$) as a function in incident velocity between one (complete penetration) and zero (complete reflection), while the later always changes smoothly as a wave nature. 
Here we report a systematic study of the quantum tunneling property for a bright soliton in ultracold atoms, which behaves as a classical particle (matter wave) in the limit of small (large) incident velocity. In the intermediate regime, the classical and quantum properties are combined via a finite (but not full) discontinuity in the tunneling transmission coefficient.
We demonstrate that the formation of a localized bound state is essential to describe such inelastic collisions, showing a nontrivial nonlinear effect on the quantum transportation of a bright soliton.
\end{abstract}


\maketitle

\begin{figure}
\center
\includegraphics[width=7.8cm]{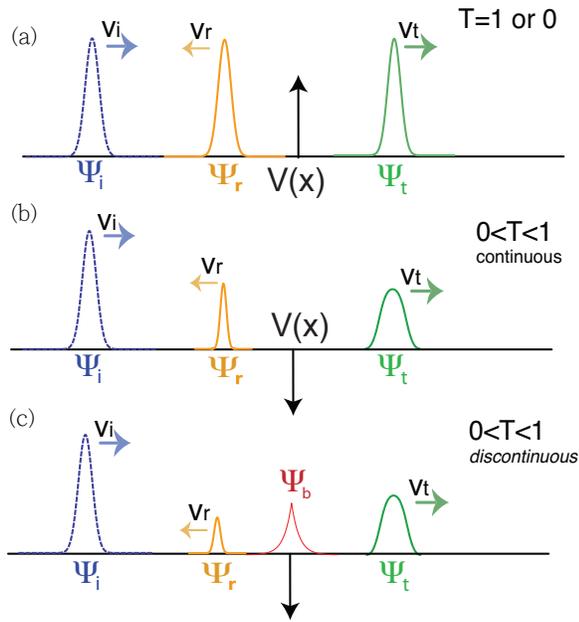}
\caption{(Color online) Schematic plots for different tunneling dynamics through a local potential, $V(x)$. $v_i$, $v_t$, and $v_r$ are the incident, transmission, and reflection velocities, respectively. (a) Classical picture with either a total transmission ($T=1$) or total reflection ($T=0$). (b) Quantum mechanical picture with a partial transmission ($0<T<1$). The notations, $\Psi_i, \Psi_t$, and $\Phi_r$ represent the incident, transmitted, and reflected wavefunctions. (c) Inelastic scattering process of a BS through a potential well ($V_0<0$), where a localized bound state, $\Phi_b$, appears after scattering.}
\label{schematic plot}
\end{figure}
The interplay between the interaction effect and the disorder potential has long been an interesting subject in condensed matter physics, from Anderson localization in the noninteracting limit \cite{anderson} to the Bose glass in the strongly interacting region \cite{Boseglass}. Similar transport problem can also be investigated in the system of ultracold atoms, where Bose-Einstein Condensates (BECs)  are demonstrated to unify concepts in classical and quantum physics at a macroscopic scale \cite{BECnew}.
Fermionic or bosonic particles with a tunable interaction strength can be studied in a well-controlled quasi-disordered potential \cite{disorder}.
In this context,  solitons,  localized wavepackets undergoing confinement owing to nonlinear effects~\cite{Kivshar}, become an ideal representative for the investigation in a macroscopic scale of the wave-particle duality which is one of the fundamental pillars in modern physics \cite{Griffiths}.
For example, a bright soliton (BS) resemble a classical particle in their collision properties \cite{soli}, and should have a complete penetration or reflection predicted by classical mechanics, see Fig.\ref{schematic plot}(a); on the other hand, due to the underlying matter wave nature, a soliton should always reveal partial penetration and reflection as predicted in the quantum mechanics, see Fig.\ref{schematic plot}(b). Therefore, it is of both interest and fundamental importance to study how the nonlinearity ({\it i.e.}, the interaction effect between bosonic atoms) can modify the quantum transportation properties of BS and the transition between these two regimes \cite{NLSjp, qreflec, resonant}. 

Apart from the existing literature on optical spatial solitons \cite{Cao:1995, Burtsev:1995, Goodman:2004, Linzon:2007,  Jisha}, in this paper we investigate quantum tunneling properties for a BS in one-dimensional (1D) bosonic atoms with different transport velocities.
Through a systematic numerical simulation, we find that BS is like a classical particle(Fig.\ref{schematic plot}(a)) when the incident velocity is small compared to the interaction energy and when the potential is repulsive, while it behaves as an ordinary matter wave in the other limit and is independent of the sign of the local potential (Fig.\ref{schematic plot}(b)).
In the intermediate regime, the nature of particle-wave dualism from BS shows a discontinuity in the transmission coefficient ($T$) as a function of the incident velocity, while the amplitude of the discontinuity is less than one, as required by a true classical particle.  
We numerically calculate the full phase diagram in such a crossover regime, and observe a qualitative difference in the scattering process between a potential barrier and a potential well: the latter case is an inelastic scattering due to the appearance of a localized bound state, see Fig.\ref{schematic plot}(c). Semi-analytical curves for such a border are derived  both for potential barriers as well as potential wells. The dual nature in quantum tunneling of BS elucidated in this work should be ready to be observed in the system of ultracold atoms as well as in the dielectric material with electromagnetic waves. 

Here, we consider the dynamics of a weakly interacting BEC at zero temperature, which can be well-approximated by the Gross-Pitaevskii  equation, referred also as the nonlinear Schr\"odinger equation (NLSE) \cite{GP, note1}, 
\begin{equation}\label{nlse}
\left[-\frac{1}{2}\frac{\partial^2}{\partial x^2}+g|\Psi(x,t)|^2+V(x)\right]\Psi(x,t)=i\frac{\partial}{\partial t }\Psi(x,t),
\end{equation}
where the particle mass $m$ and $\hbar$ are both set to $1$, $\Psi(x,t)$ represents the condensate wavefunction, $g$ measures the inter-particle interaction, and $V(x)=V_0\delta(x)$ indicates a defect potential.
When the interaction is attractive, $g<0$, a stable bright soliton is supported in a uniform system with the solution 
\begin{equation} 
\Psi_{\rm bs}(x,t)=\frac{\beta}{\sqrt{|g|}}{\rm sech}\left[\beta(x-x_c-v_i\,t)\right]e^{i\theta(x,t)},
\label{bs}
\end{equation}
where the center of the wavepacket is denoted by $x_c$, and $\theta(x,t)\equiv v_i x-E t$, with the total energy $E\equiv v_i^2/2+\mu$  and  the chemical potential $\mu=-\beta^2/2$, respectively.
The velocity for BS is characterized by $v_i$.
By taking $\beta=|g|/2$ to have  a unit normalization, {\it i.e.}, 
 $\int_{-\infty}^\infty |\Psi_{\rm bs}(x,t)|^2\,dx=2\beta/|g|=1$, the only two independent parameters left are the normalized potential strength,  $\tilde{V}_0 \equiv V_0/|g|$, and the normalized initial velocity, $\tilde{v}_i \equiv v_i/|g|$, which defines our parameter space.
In the following, we consider the transportation process when such BS wavepacket is generated at $t=0$, centered at $x_c\to-\infty$, and then propagates along the positive $x$-axis with an initial velocity ${v}_i$.
This wavepacket then scatters the local defect $V(x)$ at the position $x=0$, resulting in possible transmitted, reflected, and localized wave functions after a certain time measured.

\begin{figure}
\center
\includegraphics[width=10.0cm]{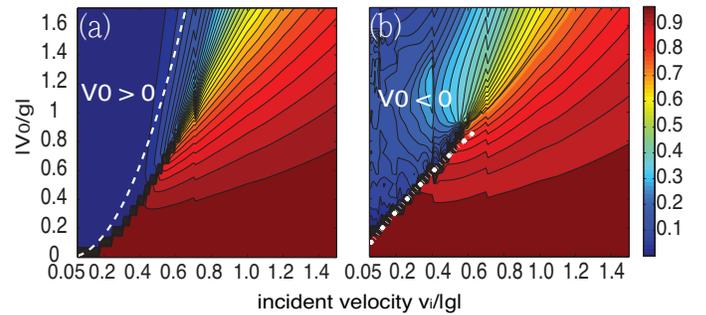}
\caption{(Color online) Contour plot of the transmission coefficient as a function of incident soliton velocity $v_i$ and potential strength $|{V}_0|$.
Cases of potential barrier ${V}_0 > 0$ and potential well ${V}_0 < 0$ are separately shown in (a) and (b). White dashed lines are plotted from Eq.\eqref{boundary_classical}  and Eq.\eqref{compare} for the corresponding analytical results (see the text). 
}
\label{phasecomp}
\end{figure}
First of all, the calculated transmission coefficient ($T$)  as a function of ${v}_i$ and ${V}_0>0$  for a repulsive potential is illustrated in Fig.\ref{phasecomp}(a) by  directly solving Eq.(1) numerically. 
 Here,  $T$ is defined as $\equiv \int_{a}^\infty \lim_{t\rightarrow\infty}|\Psi(x,t)|^2\,dx$ with a small value $a>0$ to exclude the contribution from any possible localized bound states.
As one can see from Fig.\ref{phasecomp}(a), in the region with a small value of ${v}_i$ and $|{V}_0|$, there exists a line that characterizes the discontinuity in the  transmission coefficient.
The existence of such a discontinuity  certainly reflects the particle nature of BS,  i.e. totally transmitted ($T =1$) or totally reflected ($T=0$) as shown in Fig. \ref{schematic plot}(a).
%
%
However, this line of border for the particle nature breaks down at a critical point in the parameter space where the incident velocity and corresponding potential strength are denoted by ${v}^\ast_i \sim 0.8$ and $|{V}^\ast_0|\sim 1.2$.
Beyond these values, instead of a disrupt change, the  contours of transmission coefficient $T$ changes continuously as a regular matter wave.

On the contrary, in Fig.\ref{phasecomp}(b), the tunneling properties are different for a BS through an attractive potential (${V}_0<0$, a potential well).
Qualitatively speaking, we have a similar ``phase diagram" as the case of a potential barrier, but now the {\it phase boundary} to illustrate the particle nature becomes a nearly linear one.
 As it would be demonstrated later,  at  a small value of ${V}_0$, such a universal border, independent from any additional parameters,  comes from the existence  of a localized bound state.
The formation of this localized bound state screens the potential well and results in extra interactions and nonlinearity on the quantum tunneling of BS.

In order to give a deeper understanding of these numerical results, we consider the  limit of a weak nonlinear interaction for the first step.
When the interaction effect is much smaller than the kinetic energy and potential energy, the corresponding transmission coefficient should be similar to that in the standard quantum mechanics textbook~\cite{Griffiths}, {\it i.e.}, 
\begin{equation}
T\approx\frac{\left({v_i}/{V_0}\right)^2}{1+\left({v_i}/{V_0}\right)^2}+O(\frac{1}{v_i^2}).
\label{T_regular}
\end{equation}
As expected for a characteristic wave nature, the transmission coefficient $T$ is always  continuous and independent of the sign of potential strength, $V_0$.
We note that above results are true for the incident wave as either a soliton wavepacket or a plane wave.
In such a scenario, the incident BS can be easily distorted by the local potential due to that the nonlinearity is too weak to support the original soliton solution, resulting in a lots of dispersive radiations in the transmitted or reflected waves~\cite{fast07}. 

On the other hand, in the limit of a weak and repulsive potential along with a  small velocity, {\it i.e.}, the  strong interaction limit, we can safely assume that the propagating soliton is not  affected by the potential.
Hence, one can use the center position of BS, $x_c(t)$, to describe the whole transportation process if there is no bound state generated during the scattering process.
In this limit, one can rigorously show that the dynamics of $x_c(t)$ behaves like a classical particle moving effectively in a conservative potential, $V_{\rm eff}(x)$, which is just a convolution of the local potential with the soliton wavefunction \cite{slow07}, {\it i.e.}, $V_{\rm eff}(x)=\frac{\beta^2}{2|g|}V_0\,\text{sech}^2(x)$.
Therefore, the corresponding  ``conservation law" for the total energy is found to be,
\begin{equation}
{v}(t)^2+\frac{\beta^2}{|g|^2}{V}_0\,\text{sech}^2(x_c(t))={v}_i^2+\frac{\beta^2}{|g|^2}{V}_0\,\text{sech}^2(x_i),
\label{boundary_classical}
\end{equation}
where ${v}(t)\equiv \frac{1}{|g|}\,\frac{dx_c(t)}{dt}$ is the defined particle velocity for BS.
As a result, the border across a total reflection and a total transmission can be defined by taking ${v}(t)=0$ and $x_c=0$ as the boundary condition, along with the initial condition  $x_i\to-\infty$. 
Then, we obtain the relation  $V_0/|g|=(|g|/\beta)^2(v_i/|g|)^2=4(v_i/|g|)^2$, which is depicted as the white dashed line in Fig.\ref{phasecomp}(a).

However, we know that above semi-classical approach used for  Eq.(\ref{boundary_classical})  fails when ${V}_0$ is larger than a critical value, denoted as  ${V}_0^\ast$ due to the breakdown of taking the BS as a classical particle.
This critical value can be estimated as following: as the center of BS arrives the location of a potential, it gains a local potential energy, ${V}_0|\Psi(0)|^2$, which cannot be larger than the absolute value of the chemical potential, $|\mu|=\beta^2/2=|g|^2/8$, in order to keep the soliton description valid.
From Eq.(\ref{bs}), we have the critical value for the breakdown as ${V}_0^\ast/|g|=0.5$. 
In Fig.\ref{phasecomp}(a), the agreement between the analytical curve defined by Eq.(\ref{boundary_classical}) and our direct numerical simulations is both qualitatively and quantitatively. 
When ${V}_0$ is close to ${V}_0^\ast$, the BS is in the brink of collapsing, then the reflection part as well as the non-soliton radiation become not negligible.
Discrepancies from the above analytical results are therefore expected.

Now we come to the potential well, which should have similar results as the potential barrier  in the limit of large ${V}_0$ and ${v}_i$ (the wave nature in the weak interaction limit).
However, in the regime of small ${V}_0$ and ${v}_i$, the wavepacket description used above fails for the lack in the consideration of  possibility  localized bound states supported in an attractive interaction.
The appearance of a localized bound state indicates extra inelastic scattering and, therefore, changes the resulting tunneling amplitude dramatically, as compared to the case of a potential barrier.
The bound state wavefunction for a localized potential has been well-studied in the literature \cite{delta_exact1, delta_exact2, delta_exact3, delta_exact4}, and its analytic form can be written as following:
\begin{equation}
\Psi_b(x)=\frac{\beta_b}{\sqrt{|g|}}{\rm sech}(\beta_b|x|+x_b),
\label{Psi_b}
\end{equation}
where $\beta_b$ measures the slope and amplitude of the bound state, and $x_b\equiv\tanh^{-1}(|V_0|/\beta_b)$ is the shift of effective peak position from the potential center. 
It is easy to see that the bound state wavefunction is composed of two soliton-like solutions, but with different center positions and $\beta$.
By matching the discontinuity in the wavefunction slopes with the potential strength, for a given renormalization of the bound state, {\it i.e.}, $\beta_b$ is fixed, such bound states exist only in a weak potential limit and disappears when $|{V}_0|>\beta_b$.
Such an anti-intuitive result originates from the fact that the maximum slope of BS is limited by its renormalization due to the nonlinear (interaction) effect.
Although the localized bound state also exists in a repulsive potential defect,
it cannot be easily produced in the tunneling process due to the mismatch in the boundary conditions.
Therefore it does not affect the tunneling property as we discussed above. 

Inspired by the numerical simulations of the tunneling process not shown her), we consider the following simplified picture of the tunneling dynamics in the presence of a bound state, {\it i.e.}, inelastic scattering.
To derive an analytical formula for the border when a soliton scatters by a potential well, we assume: (i) the reflected wavefunction is negligible, (ii) the bound state appears after the scattering, and (iii) the transmitted wave also has a soliton profile.
Since both the soliton solution and the localized bound state are governed by two parameters, $\beta$ and ${v}$, as shown in Eq.(\ref{bs}), the relevant parameters to describe a soliton tunneling are therefore: $(\beta_i, {v}_i)$ for the incident soliton, $(\beta_t, {v}_t)$ for the transmitted one, and $(\beta_b, {v}_b=0)$ for the localized bound state.
Since ${v}_i$ is given and $\beta_i\equiv |g|/2$ is required for the initial unit normalization, now we have only three parameters to be determined: $\beta_b$, $\beta_t$, and $v_t$. 

Instead of matching the boundary of wavefunctions during the scattering process, we use the conservation laws to extract the three unknown parameters above in a more general method. 
Based on the  above three assumptions, we can write down three equations:
\begin{eqnarray}
\frac{2\beta_i}{|g|}&=& 1 = \frac{2}{|g|}(\beta_b-|V_0|)+T,
\label{prob} \\ 
-\frac{\beta_i^2}{2}+\frac{v_i^2}{2}&=&
\frac{-\beta_b^2}{2}(1-T)
+\left(\frac{-\beta_t^2}{2}+\frac{v_t^2}{2}\right)T, \label{energy}\\
v_i&=&\frac{2\beta_t}{|g|}v_t=Tv_t
\label{momentum}
\end{eqnarray}
where we have expressed the transmission coefficient $2\beta_t/|g|$ as $T$ via assumption (iii).
Equations (\ref{prob}) and (\ref{energy}) represent the conservations of total probability and total energy by including both the localized bound state and the transmitted soliton.
Equation (\ref{momentum}) can be understood as the conservation of current density due to the change of soliton amplitude. We note that $|V_0|<\beta_b$ in order to have $T>0$ in Eq.(\ref{prob}). 
By eliminating the other two variables, one can obtain $T$ as a function of the normalized potential strength $\tilde{V}_0\equiv |V_0/g|$ and the normalized initial velocity $\tilde{v}_i\equiv v_i/|g|$,
\begin{eqnarray}
T=\frac{\tlV_0(\tlV_0+1)\mp\sqrt{\tlV_0^2(\tlV_0+1)^2-(4\tlv_i^2\tlV_0+3\tlv_i^2)}}{2\tlV_0+3/2},
\label{compare}
\end{eqnarray}
where the ``+" solution is physically invalid. 
From Eq.(\ref{compare}), we find several interesting properties in the tunneling of BS through a potential well.
First of all, above solution for the transmission coefficient is real only when $|\tlV_0|^2(|\tlV_0|+1)^2-(4\tlv_i^2|\tlV_0|+3\tlv_i^2)\geq 0$, or when ${v}_i$ is smaller than a critical velocity, ${v}_c(|V_0|)$, {\it i.e.},
\begin{eqnarray}
\frac{v_c}{|g|}\equiv \frac{|V_0/g|\left(|V_0/g|+1\right)}{\sqrt{3+4|V_0/g|}},
\label{v_c}
\end{eqnarray}
where ${V}_0$ and ${v}_i$ has to be bounded by requiring $T<1$.
In  Fig.\ref{phasecomp}(b), the border for the discontinuity in the transmission coefficient is compared with our analytical formula and direct numerical simulations, which results in  very good agreement.
More importantly, we find that the critical velocity defined in Eq.(\ref{v_c}), {\it i.e.}, the curve for the discontinuous transmission coefficient, shows a rather straight line (although is not exact),  instead of a parabolic one for the potential barrier.
The major reason is due to the existence of a localized bound state in this inelastic scattering process.


Although our simplified theory does predict the locations of transition, the transmission coefficient $T$ we obtained in Eq.(\ref{compare}) is slightly less than the values obtained by direct numerical simulations. This is not surprising because we have neglected the radiation parts (non-soliton waves) in the transmitted waves, which cannot be captured in such a simple theory.
Last but not least, we restate that the estimations we made here for the tunneling dynamics of a potential well  cannot be applied to the potential barrier, because the bound state wavefunction supported by a potential barrier has to a double-humped one in the profile due to a mismatching boundary condition. 
On the side of  wave nature, the transmission coefficient is identical both for potential well and potential barrier when one  considers a regular (noninteracting) matter wave tunneling, but on the side of particle nature, it turns out to be very different in the classical particle picture. This also reflects the non-trivial effect of interaction (nonlinearity) in the soliton scattering problem.

In conclusion, we have systematically studied the tunneling of a bright soliton  subjecting to a localized potential defect.
By performing direct numerical simulations, we obtain a full phase diagram of the transmission coefficient in terms of the incident velocity and potential strength. Our results show an fundamentally important transport property, which can be easily observed in ultracold atoms, nonlinear optics, or even soft-matter systems.

This research is supported by the National Science Council in Taiwan.



\begin{thebibliography}{999}
\bibitem{note1}
We note that such a mean-field approximation in principle fails in 1D uniform system, where the quantum fluctuation can destroy the condensate particles. However, for a system of finite length and of a large number of atoms, this mean-field approximation can still be a very good approximation to understand the collective behavior. Besides, similar NLSE can be also used to describe the  EM wave propagating in a nonlinear media, where no quantum fluctuation correction is needed in the ordinary field strength.
\end{thebibliography}
\end{document}